\documentclass[%
superscriptaddress,
 preprint,
showpacs,preprintnumbers,
amsmath,amssymb,
aps,
prb,
]{revtex4-1}

\usepackage{graphicx}% Include figure files
\usepackage{dcolumn}% Align table columns on decimal point
\usepackage{bm}% bold math
\usepackage{upgreek}
\usepackage{amsmath}

\begin{document}

%\preprint{APS/123-QED}

\title{Single-sized phase-change metasurfaces for dynamic information multiplexing and encryption}% Force line breaks with \\

\author{Tingting Liu}
\affiliation{Institute for Advanced Study, Nanchang University, Nanchang 330031, China}
%\affiliation{Jiangxi Key Laboratory for Microscale Interdisciplinary Study, Nanchang University, Nanchang 330031, China}

\author{Jie Li}
\affiliation{Optoelectronic Sensor Devices and Systems Key Laboratory of Sichuan Provincial University, College of Optoelectronic Engineering, Chengdu University of Information Technology, Chengdu 610225, China}

\author{Shuyuan Xiao}
\email{syxiao@ncu.edu.cn}
\affiliation{Institute for Advanced Study, Nanchang University, Nanchang 330031, China}
%\affiliation{Jiangxi Key Laboratory for Microscale Interdisciplinary Study, Nanchang University, Nanchang 330031, China}

\begin{abstract}
Optical metasurfaces empower us to manipulate the electromagnetic space and control light propagation at the nanoscale, offering a powerful tool to achieve modulation of light for information processing and storage. In this work, we propose a phase-change metasurface to realize dynamic multiplexing and encryption of near-field information. Based on the orientation degeneracy and polarization control governed by Malus’s law, we elaborately design the orientation distribution of Sb$_2$S$_3$ meta-atoms with the same dimension to simultaneously satisfy the amplitude modulation requirements of different channels. Using the corresponding polarization control as decoding keys, three different nanoprinting images can be displayed, and these multiplexed images can be switched on and off by leveraging the reversible tunability of the Sb$_2$S$_3$ meta-atoms between the amorphous and crystalline states. With the unparalleled advantages of ultra-compactness, simple design strategy, high information density and security, the proposed metasurfaces afford promising prospects for high-end applications in ultracompact and intelligent dynamic display, high-dense optical data storage, optical information encryption, etc. 

\end{abstract}

%\pacs{42.70.-a, 42.79.-e, 78.67.Pt}% PACS, the Physics and Astronomy
\maketitle

%\tableofcontents

\section{\label{sec1}Introduction}

The ability to modulate light on demand is the key to modern photonic systems. Recent decade has witnessed remarkable progress in metasurfaces which enable optical elements with potential size, weight, power, and cost benefits, especially the flexibility to tailor intrinsic properties of incident light such as amplitude, phase, and polarization at the subwavelength \cite{yu2011light,sun2012gradient,Kildishev2013,Kamali2018,Neshev2018}. By carefully designing planar nanostructures with spatially varying field distributions across the surface, metasurfaces have led the way to meta-optics for many optical elements with various functionalities such as beam steering \cite{Wan2021,Thureja2020}, focusing \cite{Khorasaninejad2016, Lin2019, Jin2020, Li2021, Li2022, Arbabi2022, Huang2023}, spin Hall effect \cite{Ling2017, Li2018, Chaudhuri2018, Zhu2020, Kim2022}, holographic \cite{Huang2013, Ren2020, Bao2021, Liu2021a, Xiong2023} and nanoprinting \cite{Li2020three, Deng2020, Fu2022, Dai2022} imaging. Benefiting from the design flexibility, metasurfaces-based devices are evolving from single to multiple functionalities, exhibiting significant potential in information storage and encryption. Previous researches have reported different design approaches such as segmenting \cite{Wei2019, Chen2020}, layer stacking \cite{Zhou2018, Hu2019, Yao2020, Georgi2021, Malek2022}, and interleaving \cite{Wang2016, Deng2020a, Du2021, Cao2022, Yuan2023}, for multi-channel image displaying and information encoding. Yet most of them are essentially the simple hybrids of several single-functional devices, since the operational zones of metasurfaces are divided into several segments corresponding to different functionalities. In order to further enhance information density and security, single-cell metasurfaces by taking full advantages of the design degrees of freedom of a single nanostructure are widely explored \cite{BalthasarMueller2017, Jin2018, Hu2019a, Dai2020, Mehmood2022}. Nevertheless, the integration of multifold information usually requires many nanostructures with different dimensions patterned in a single piece, inevitably increasing the fabrication difficulty. Therefore, there is a strong need to develop a simple design strategy using single-sized metasurfaces enabling high level of integration and miniaturization in date storage and encryption.

Most recently, metasurfaces are in the revolutionary process from passive to active tunability towards intelligent integrated photonic devices. The active metasurface with dynamical optical responses is highly desirable in realizing agilely switchable and reconfigurable photonic functionalities, allowing new opportunities for tailoring the light propagation and interaction with matter. Because the optical response of metasurface usually relies on the dimensions and dielectric properties, substantial efforts have explored the tuning mechanisms of integrating active materials into nanostructures. By leveraging external stimuli such as mechanical actuations, chemical reactions, optical, electrical and thermal schemes, light field distributions of these active metasurfaces exhibit dynamically controllable functionalities, offering a programmable flexibility in information processing and storage\cite{Xiao2020}. In particular, chalcogenide phase-change materials (PCMs) are uniquely poised for the photonic modulation and resonance tuning of active metasurfaces, owing to their striking portfolio of properties \cite{Wuttig2017, Ding2019, Choi2019, Leitis2020, Liu2021, Abdollahramezani2022, Li2023}. PCMs can be rapidly and reversibly switched between amorphous and crystalline states, and the two states show pronounced contrast in optical and electronic properties. In recent studies, Sb$_2$S$_3$, Sb$_2$Se$_3$, and GeSe$_3$ have been identified as a family of highly promising ultralow-loss PCMs for nanophotonic devices over the visible and mid-infrared spectrum \cite{Delaney2020}. They have been successfully demonstrated in various reconfigurable metasurface devices for information storage and display such as high-resolution color \cite{Lu2021, Luo2023, liu2023non}, beam steering \cite{Qin2021, Chen2021}, and holographic display \cite{Moitra2022, Liu2022}. However, the realization for dynamic information multiplexing and encryption via single-sized phase-change metasurfaces remains unexplored. 

In this work, we propose a simple design approach which enables dynamic intensity modulation for switchable three-channel nanoprinting imaging in a single-sized metasurface. Through elaborately controlling the orientation angles of an anisotropic Sb$_2$S$_3$ meta-atom governed by Malus’s law, it creates the degeneracy of energy allocation in different channels and achieves nanoprinting information multiplexing in a single metasurface, allowing for metasurface encryption under the combinations of polarization control. Different from the previous polarization multiplexing metasurfaces only accessible in two orthogonal-polarization states, the designed phase-change metasurface utilizes non-orthogonal polarization states for three independent information channels. The Sb$_2$S$_3$ amorphous-crystal phase transition provides active tunablity of the three-channel nanoprinting imaging from switch on to off state, further improving the information security. In the design, by simply arranging the orientations of single-sized Sb$_2$S$_3$ meta-atoms on a single-layered metasurface, the three-channel nanoprinting images can be simultaneously recorded and dynamically switchable for the first time. Furthermore, the proposed metasurface shows the advantage of manifold information and multifold encryption, offering promising prospects for applications in high-secure and high-density optical storage, ultracompact and intelligent dynamic display, optical information encryption, etc. 

\section{\label{sec2}Design Principle}

\begin{figure*}[htbp]
	\centering
	\includegraphics[width=\linewidth]{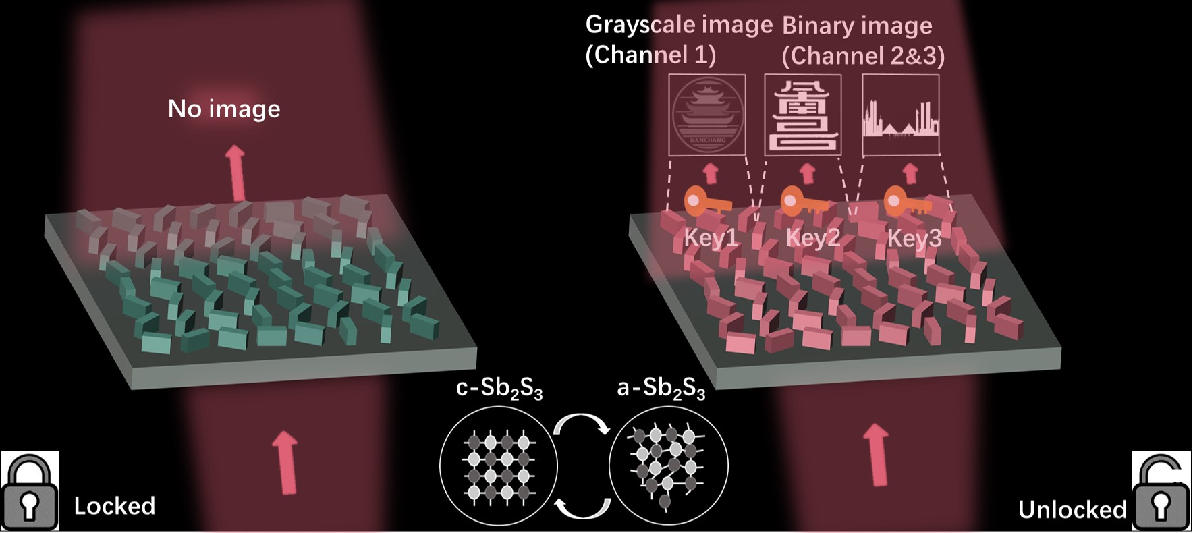}
	\caption{Schematic diagrams of the designed single-sized metasurfaces based on Sb$_2$S$_3$ for dynamic three-channel nanoprinting displays. The tunable metasurfaces exhibit three different nanoprinting images using the polarization controls as the decoding keys under the amorphous states of Sb$_2$S$_3$ (a-Sb$_2$S$_3$) and hide the image information under the crystalline states of Sb$_2$S$_3$ (c-Sb$_2$S$_3$). The on and off modes can be switched reversibly between the two states.  }
	\label{fig1}
\end{figure*}

Figure 1 schematically illustrates the designed phase-change metasurface enabling dynamic control of multichannel nanoprinting displays by adjusting the amorphous-crystal phase transition states of Sb$_2$S$_3$. The image information of three independent channels can be encoded into a single-sized phase-change metasurface, through carefully arranging the Sb$_2$S$_3$ nanobricks with fixed size but different orientations to modulate light intensity pixel by pixel. In the amorphous state, the metasurface can be characterized by the amplitude profiles that encode the near-field information. Using corresponding polarization controls as the decoding keys, the three different nanoprinting images can be displayed at the transmission side. After Sb$_2$S$_3$ crystallizes from an amorphous state into an orthorhombic structure of fully crystalline state by heating, the meta-atoms exhibit a substantial change in refractive index, and thus the image information encoded by amplitude modulation recorded on the metasurface cannot be read by arbitrary polarization keys. As a result, the nanoprinting display are switched into off state in which the coded images are hided. The switchable display driven by phase-change meta-atoms affords another degree of freedom in information encryption.

In the phase-change metasurface structure for nanoprinting display, each unit cell is composed of a Sb$_2$S$_3$ nanobrick with fixed length $L$, width $W$, and height $H$ sitting on a planar substrate SiO$_2$, as shown in Fig. 2(a). For practical fabrication and modulation, the nonreactive dielectric material, a Si$_3$N$_4$ film with a thickness of 70 nm is employed to encapsulate the Sb$_2$S$_3$ nanobricks \cite{Lu2021, Moitra2022}. Such configuration can create an undisturbed environment for stable performance of the nanostructures, and protect the chalcogenide material from the heat damage during the amorphous-crystal phase transition process such as avoiding sulfur loss through evaporation in thermal switching. Here we select Sb$_2$S$_3$ as a nonvolatile PCM to construct the metasurface operating in the visible spectrum. Compared to the most considered PCMs such as GST and VO$_2$, Sb$_2$S$_3$ has a wide bandgap of 1.70-2.05 eV, leading to the absorptance band-edge moving to the visible spectrum around 600 nm \cite{Dong2018}. As shown in Fig. 2(b), the refractive index of Sb$_2$S$_3$ keeps high around 3.5 in the visible spectrum, being an excellent counterpart to widely-used Si$_3$N$_4$ and TiO$_2$ for visible all-dielectric metasurfaces. The refractive index contrast between the amorphous and crystalline states is around 0.5 in both the real and imaginary component at the wavelength of interest $\lambda$=633 nm, allowing dynamic switching of the optical properties. 
%Using the approximate applied energy and stimulus duration on Sb$_2$S$_3$, the density of Sb$_2$S$_3$ increases upon crystallization, leading to the changes in optical properties of the material. The effective dielectric constant of Sb$_2$S$_3$ can be modeled by Lorentz-Lorenz relation as \cite{Tian2019, Meng2021, Liu2022}
%\begin{equation}
	%\frac{\varepsilon_{\text{eff}}-1}{\varepsilon_{\text{eff}}+2}
	%=m\frac{\varepsilon_{c}-1}{\varepsilon_{c}+2}
	%+(1-m)\frac{\varepsilon_{a}-1}{\varepsilon_{a}+2},
	%\label{eq1}
%\end{equation}
%where $\varepsilon_{a}$ and $\varepsilon_{c}$ are the permittivity of a-Sb$_2$S$_3$ and c-Sb$_2$S$_3$, respectively, $m$ represents the crystalline fraction with a value ranging from 0 to 1 for amorphous to crystalline state.

\begin{figure}[htbp]
	\centering
	\includegraphics
	[width=\linewidth]{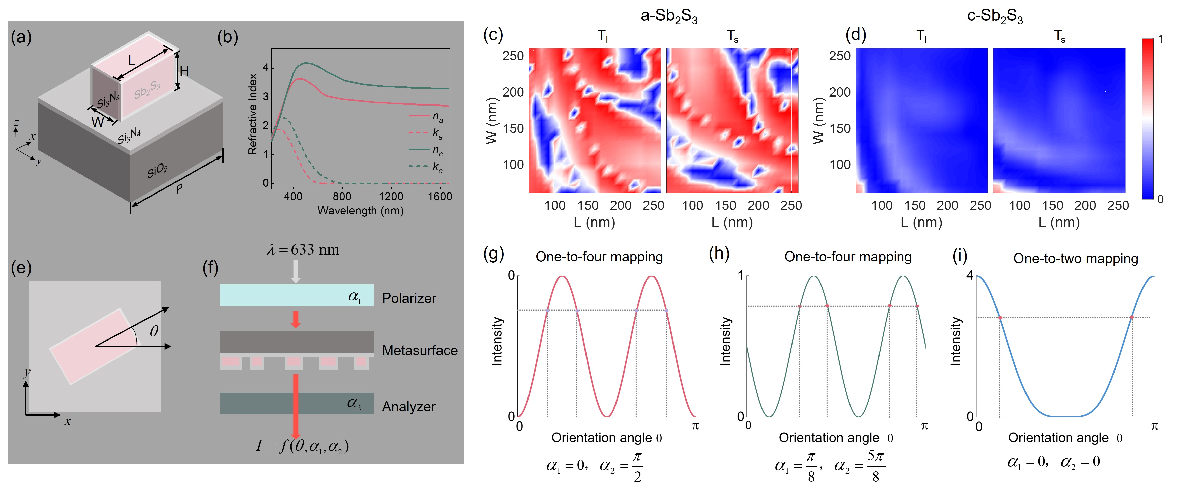}
	%[scale=0.8]    
	\caption{Unit-cell structure and design principle. (a) Schematic illustration of a unit cell of the metasurfaces with defined structure parameters. (b) The refractive index of Sb$_2$S$_3$ for both amorphous and crystalline states. (c, d) Simulated transmission with incident light polarized along the long and short axes of the nanobrick of a-Sb$_2$S$_3$ and c-Sb$_2$S$_3$ metasurfaces, respectively. (e) Planar illustration of a unit cell with a rotation angle $\theta$. (f) Principle of intensity modulation in near-field nanoprinting displays. When the metasurfaces are inserted between a polarizer and analyzer, the transmission intensity varies with the rotation angle of the nanobrick $\theta$, the polarization directions of polarizer and analyzer, i.e. $\alpha_{1}$ and $\alpha_{2}$. (g-i) The transmission intensity as a function of $\theta$ with $I_{0}=4$, as the polarization direction combinations of polarizer and analyzer are (0, $\pi/2$), ($\pi/8$, $5\pi/8$), and $(0,0)$, respectively. }
	\label{fig2}
\end{figure}

To create the encoding-freedom for light polarization, the phase-change metasurface is inserted between a bulk-optic polarizer and an analyzer. As shown in Fig. 2(f), the incident light passes through a polarizer with polarization direction $\alpha_{1}$, a nanobrick with rotation angle $\theta$, and an analyzer with polarization direction $\alpha_{2}$. According to Malus’s law, a general expression of the amplitude (intensity) and polarization state of the output light can be written as 
\begin{equation}
	I=I_{0}|\frac{t_l+t_s}{2} \cos(\alpha_{2}-\alpha_{1})+\frac{t_l-t_s}{2}\cos(2\theta-\alpha_{2}-\alpha_{1})|^2,
   \label{eq2}
\end{equation}
where $I_{0}$ is the light intensity after the polarizer, $t_l$ and $t_s$ are the complex transmission coefficients along the long and short axis of the nanobrick, respectively. The derivation of Eq. (\ref{eq2}) can be found in the Section 1 of Supplementary Information. The output transmission amplitude can be regarded as a function dependent on the polarization direction $\alpha_{1}$, $\alpha_{2}$ and nanobrick orientation $\theta$, and thus the amplitude modulation can be realized by carefully setting these polarization combination and meta-atom orientation. 

If the incident light is linearly polarized along $x$ axis after polarizer with $\alpha_{1}=0$, and the output transmission light through the analyzer with $\alpha_{2}=\pi/2$ is along the $y$ axis, the amplitude in Eq. (\ref{eq2}) can be calculated in a simple form as 
\begin{equation}
	I_{1}=I_{0}|\frac{t_l-t_s}{2}|^{2} \sin^{2}(2\theta).
   \label{eq3}
\end{equation}
For the designed anisotropic nanobrick in Fig. 2(a), $|\frac{t_l-t_s}{2}|^{2}$ is a constant that only depends on the phase difference between the long and short axis, which is usually considered as the cross-polarization conversion efficiency of the anisotropic nanobrick. Here the transmission amplitude along the long and short axes are set as $T_l=1$ and $T_s=0$, respectively. As a result, by adjusting the orientation $\theta$ of each meta-atom, the near-field amplitude can obtain continuous manipulation, and then a continuous grayscale image can be encoded into the metasurface pixel by pixel. This can be observed using the normalized curve depicted in Fig. 2(g). In particular, there is more than one choice of the orientation angle, including $\theta$, $\pi/2-\theta$, $\pi/2+\theta$, and $\pi-\theta$, to generate a specific output amplitude. Such orientation degeneracy offers degrees of freedom in designing meta-atoms to establish information channel. In the second near-field channel, we set $\alpha_{1}=\pi/8$ and $\alpha_{2}=5\pi/8$ respectively, written in the form of $(\pi/8, 5\pi/8)$. The output transmission amplitude in (\ref{eq2}) can be calculated as 
\begin{equation}
	I=I_{0}|\frac{t_l-t_s}{2}|^{2} \sin^{2}(2\theta-\pi/4).
	\label{eq4}
\end{equation}
As depicted in Fig. 2(h), the orientation degeneracy with one-to-four mapping still works for such polarization control. Then we introduce the third channel by setting $\alpha_{1}=0$ and $\alpha_{2}=0$ with the transmission amplitude calculated as 
\begin{equation}
	I=I_{0}|\frac{t_l-t_s}{2} \sin(2\theta)+ \frac{t_l+t_s}{2}|^{2}.
	\label{eq5}
\end{equation}
With $T_l=1$ and $T_s=0$, it can be simplified as $I=\frac{I_{0}}{4}(\sin(2\theta)+ 1)^{2}$. As depicted in Fig. 2(i), it can be observed that the amplitude curve shows a one-to-two mapping degeneracy with quite different tendency compared with those in Fig. 2(g) and (h). Such polarization combination of polarizer and analyzer brings another way to modulate amplitude in the new channel. 

The three equations form the basis for realizing multifold integration and dynamic display of nanoprinting images with a single-sized phase-change metasurface design approach. To be exact, the fundamental of this approach lies in intelligently engineering the amplitude of incident waves under polarization controls. Based on the degeneracy of energy allocation governed by Malus’s law, the multiplexing information channel can be established by varying nanobrick orientation and controlled by polarization combinations. Meanwhile, dynamically tunable display can be realized by adjusting crystalline state of Sb$_2$S$_3$ meta-atoms under the thermal, electrical, or optical external stimulus. It is interesting that the three channels are controlled by the non-orthogonal polarization states, i.e. the combinations of $(0,\pi/2)$, $(\pi/8, 5\pi/8)$, and $(0,0)$ of polarizer and analyzer in the optical path, which is different from the usual case in which anisotropic nanostructures exhibit distinct optical properties in the two orthogonal polarization states. Such non-orthogonal polarization multiplexing metasurface takes full advantage of the orientation degeneracy of Malus’s law and employs polarization as the decrypted keys to improve the information capacity and security.

\section{\label{sec3}Results and discussion}

To achieve the best polarization-controlled energy allocation efficiency, the physical dimensions of the Sb$_2$S$_3$ meta-atoms are optimized at the operating wavelength. The period of the unit cell is selected as $P=400$ nm. The physical parameters including length and width of the nanobrick is optimized by keeping the period and the height $H=550$ nm fixed. The simulated transmission efficiencies along the long and short axis are depicted in Figs. 2(c) and (d) for the a-Sb$_2$S$_3$ and c-Sb$_2$S$_3$, using the finite-difference-time-domain (FDTD) method. As a result, the designed Sb$_2$S$_3$ nanobrick is 170 nm in length and 250 nm in width, with the maximum transmission along the nanobrick’s long-axis direction ($\sim$ 98$\%$) and the minimum transmission along the short-axis direction ($\sim$ 0.1$\%$) for a-Sb$_2$S$_3$. At the same time, both the transmissions for c-Sb$_2$S$_3$ are compressed to be a very low value (near zero), fulfilling the dynamic display of switching on and off.

To verify our proposed design approach, we firstly consider a simple case in which the phase-change metasurfaces achieve dynamic display of the two-channel nanoprinting images in the Section 2 of Supplementary Information. Furthermore, taking fully advantage of orientation degeneracy for amplitude responses, we are able to independently encode the three-channel nanoprinting display into a single-sized phase-change metasurface.  Fig. 3 illustrates the metasurface design flow for synchronizing display of three different images. Here the intensity modulation based on Eqs. (\ref{eq3})-(\ref{eq5}) is adopted corresponding to channel 1, 2, and 3, respectively. The continuous grayscale image of Tengwang Pavilion which is one of the most famous tower in China, is encoded in channel 1, while the binary image of Nanchang city is encoded in channel 2, and the binary image of buildings besides Ganjiang River encoded in channel 3. The amplitude profiles for $40\times40$ $\mu$m$^{2}$ metasurface are numerically calculated. The key of multiplexed amplitude modulation is the nanobrick orientation distribution of the single-sized metasurface that simultaneously satisfies the requirements for the three-channel displays. From the working principle, we elaborate the orientation encoding strategy of the three-channel images in Table 1. Based on the amplitude modulation function of $I_{1}$, $I_{2}$, and $I_{3}$, the orientation angles $\theta$ are divided into four sections. In each section, $I_{1}$ can achieve a continuous amplitude modulation ranging from $0\sim1$. In sections 1 and 3, $I_{2}$ obtains amplitude modulation lower than 0.5, while it is higher than 0.5 in sections 2 and 4. For convenience we take the lower and higher values as the binary code of ‘0’ and ‘1’ respectively. In a similar way, $I_{3}$ with lower amplitude in section 2 and 3 can be considered as the ‘0’, while higher amplitude in section 1 and 4 as ‘1’. With these intuitive relations resulting from the orientation degeneracy, the orientation of each nanobrick can be determined to meet the requirement for all the three channels. As a result, a continuous grayscale image and two binary images can be encoded into the same metasurface but displayed in three information channels under corresponding polarization control. 

The design flow is detailedly described in Fig. 3. With the amplitude profile of the selected continuous grayscale image of Tengwang Pavilion, the four orientation candidates can be obtained for $I_{1}$ using Eq. (\ref{eq3}) for Malus’s law degeneracy. To program the second channel information, the amplitude of the target image which is initially binarized is calculated for selecting the nanobrick orientations. According to the look-up relations in Table 1, two of the candidates can be determined for a lower value of '0' or a higher value of '1'. Eventually, owing to the one-to-two mapping relations of $I_{3}$ amplitude modulation, only one of the orientation candidates is chosen to further satisfy the amplitude encoding of the target binary images in the third channel. In this way, the orientations of nanostructures can be determined pixel by pixel by combining the three-channel amplitude modulation for target images, and the phase-change metasurface is designed to meet the amplitude modulation requirement for the three channels.

\begin{figure}[htbp]
	\centering
	\includegraphics[width=\linewidth]{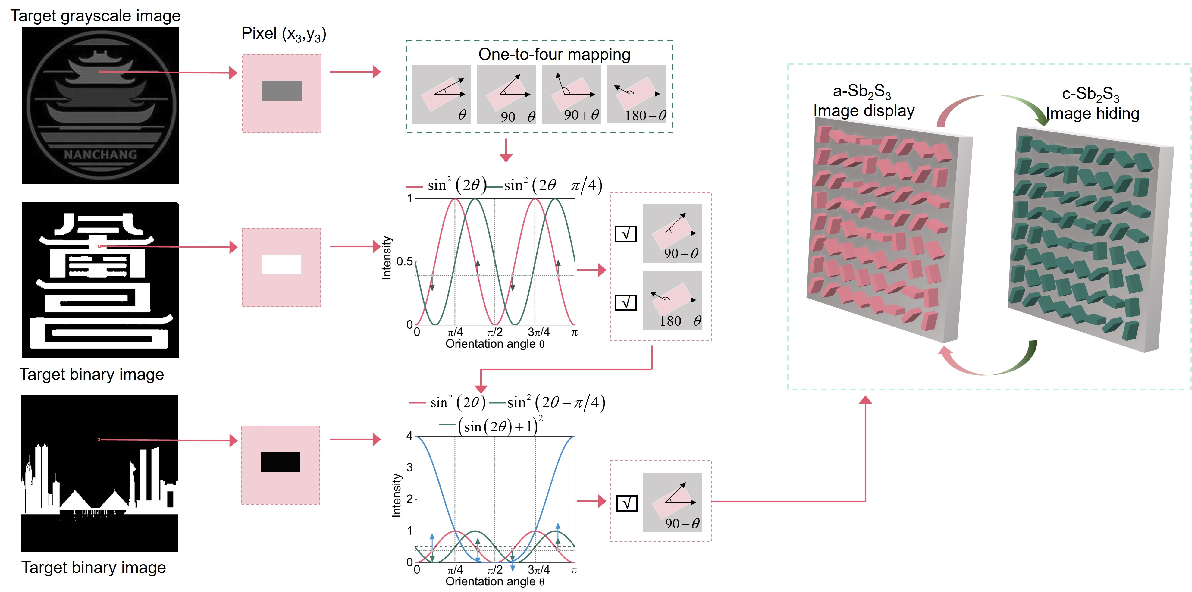}
	\caption{ Flowchart of designing the proposed three-channel phase-change metasurface. One continuous grayscale image and two binary images are chosen to encode the orientation distributions into a single-sized metasurface. Owing to the orientation degeneracy, the unit-cell amplitude of the first channel corresponds four orientations. Then two of them are selected to encode the amplitude of the second channel. At last, only one of the orientations can be determined to meet the requirement for the third channel. After the determination of the orientation distribution, three-channel metasurface can be obtained.}
	\label{Fig5}
\end{figure}

\begin{table}[ht!]
	\centering
	\caption{\bf Transmission amplitude under the polarization direction combinations of polarizer and analyzer of ($0$, $\pi/2$), ($\pi/8$, $5\pi/8$), and $(0,0)$, with the incident light intensity $I_{0}$ setting as 4.}
	\begin{tabular}{p{80pt}p{100pt}p{120pt}p{120pt}}%{cccccc}
		\hline
		$\theta$  Sections & $I_{1}=\sin^{2}(2\theta)$ &$I_{2}=\sin^{2}(2\theta-\pi/4)$ &$I_{3}=(\sin(2\theta)+ 1)^{2}$ \\
        \hline
		1: [0, $\pi/4$]         &  [0,1] & [0,0.5]:  '0' & [1,4]:  '1'\\
		\hline
		2: [$\pi/4$, $\pi/2$]   &  [0,1] & [0.5,1]:  '1' & [0,1]:  '0'\\
		\hline
		3: [$\pi/4$, $3\pi/4$]  &  [0,1] & [0,0.5]:  '0' & [0,1]:  '0'\\
		\hline
		4: [$3\pi/4$, $\pi$]    &  [0,1] & [0.5,1]:  '1' & [1,4]:  '1'\\
		\hline
	\end{tabular}
	%\label{tab:shape-functions}
\end{table}

\begin{figure}[htbp]
	\centering
	\includegraphics
	[width=\linewidth]
	%[scale=0.6]
	{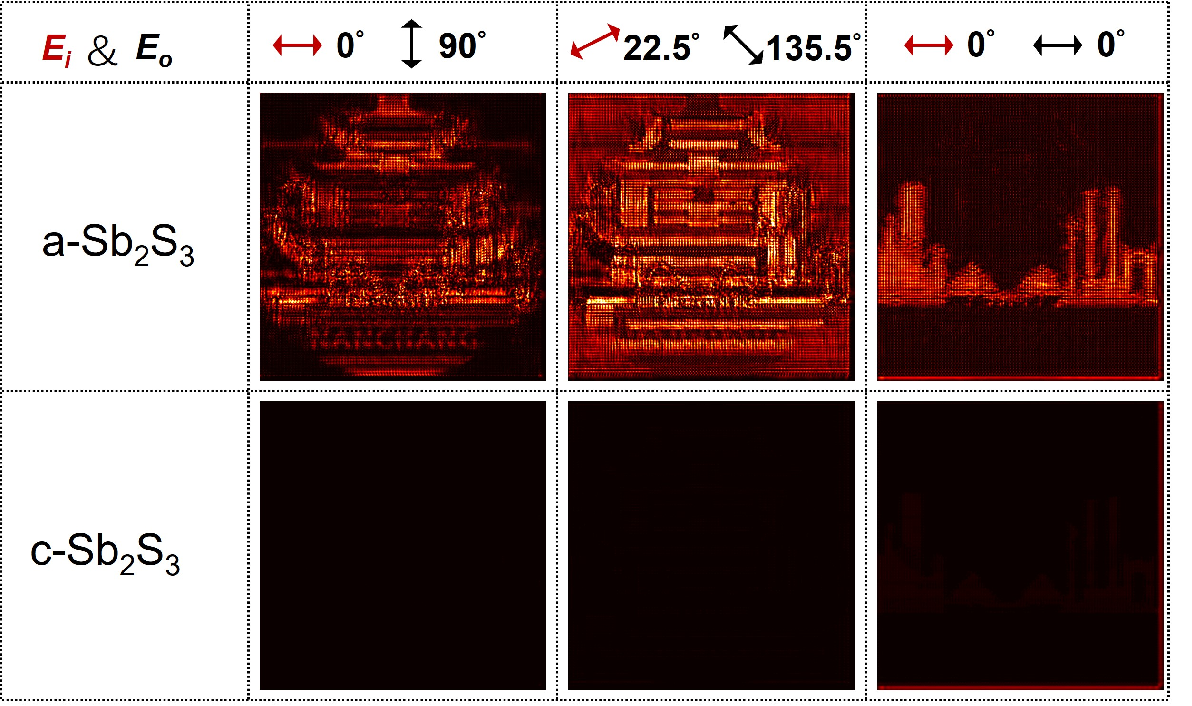}
	\caption{Simulated near-field results of the three-channel metasurface. Using the corresponding polarization combinations as the decoding keys, three nanoprinting images of the designed metasurfaces are depicted for a-Sb$_2$S$_3$ and c-Sb$_2$S$_3$ metasurfaces.}
	\label{fig6}
\end{figure}

Based on the design scheme above, the three different grayscale images are encoded into a single-sized phase-change metasurface. The metasurface composed of $100\times100$ unit cells is simulated using FDTD method at the operating wavelength at 633 nm. As shown in Fig. 4, the designed single-sized metasurface realizes the three-channel and the dynamically tunable nanoprinting displays. When the designed metasurface is inserted between the polarizer and analyzer, different images in the three channels can only be observed clearly under the preset polarization combinations. By setting the polarization directions of polarizer and analyzer as (0, $\pi/2$), ($\pi/8$, $5\pi/8$), and (0,0) respectively, the three-channel images including a continuous grayscale image of Tengwang Pavilion, a binary image of Nanchang city, and a binary image of buildings are observed in a-Sb$_2$S$_3$ metasurface, as depicted in the first row of Fig. 4. The decoded images are clear with high resolution and fidelity, implying the three-channel amplitude modulation works well with the design scheme. Meanwhile, because the light amplitude for binary images in channel 2 and 3 can only be modulated as higher or lower values, not the exactly 1 or 0, the continuous grayscale image of Tengwang Pavilion can be hazily observed as the shallow background in the two channels. Subsequently, we successfully realize the dynamic switch of nanoprinting display from on to off mode via changing the crystalline state of Sb$_2$S$_3$ metasurface, as depicted in the second row. As the Sb$_2$S$_3$ is phase transitioned to fully crystalline state, the transmission efficiency is largely varied with two orders of magnitude reduction, due to the refractive index contrast of $\Delta n \sim$0.8 and $\Delta k \sim$0.3. Such significant modulation depth for amplitude profile suggests large switch ratio of the three-channel images. Hence the preset images are hided for c-Sb$_2$S$_3$ metasurface. Specially, Sb$_2$S$_3$ can be easily be amorphized and reversibly switched on a nanosecond time scale between its amorphous and crystalline states. Benefiting from these characteristics, the phase-change metasurface can dynamically and reversibly switch the image displays between the on and off mode, implementing the image information display and encryption during the process.

\section{\label{sec4}Conclusions}

In conclusion, we have proposed and demonstrated a simple design strategy for a phase-change metasurface enabling dynamic three-channel amplitude modulation. Unlike previous multiplexing approaches such as the multi-layer, super-cell approaches, and the single-cell nanostructures with different meta-atom dimensions, the proposed strategy employs a single-sized Sb$_2$S$_3$ meta-atom to encode multiple information on the metasurface interface. By combining the orientation degeneracy of the amplitude modulation with polarization control, the proposed design strategy provides great freedom to independently manipulate each near-field channel for information encoding. As a consequence, three different nanoprinting images can be displayed using the corresponding polarization control as decoding keys. Leveraging the tunable optical properties of Sb$_2$S$_3$, the multiplexed information channels can be switched between on and off mode. Such phase-change metasurfaces provide a unique solution to realize dynamic and multifold information multiplexing and encryption without decreasing the image resolution and burdening the nanostructure design and fabrication. With the unparalleled advantages such as ultra-compactness, simple design and fabrication technique, high information density, and high resolution display, the proposed metasurfaces suggest significant potential in many applications like high-dense optical data storage, ultra-compact image displays, information encryption/security.

\begin{acknowledgments}	
	
This work was supported by the National Natural Science Foundation of China (Grants No. 12364045, 12264028, and 12304420), the Natural Science Foundation of Jiangxi Province (Grants No. 20232BAB201040 and 20232BAB211025), the Sichuan Science and Technology Program (Grant No. 2023ZYD0020), the Chenguang Program of Shanghai Education Development Foundation and Shanghai Municipal Education Commission (Grant No. 21CGA55), the Young Elite Scientists Sponsorship Program by JXAST (Grant No. 2023QT11), and the China Scholarship Council (Grant No. 202008420045). 

\end{acknowledgments}

%\bibliography{ref}% Produces the bibliography via BibTeX.

\end{document}